\PassOptionsToPackage{unicode}{hyperref}
\PassOptionsToPackage{hyphens}{url}
\documentclass[
]{article}
\usepackage{amsmath,amssymb}
\usepackage{iftex}
\ifPDFTeX
  \usepackage[T1]{fontenc}
  \usepackage[utf8]{inputenc}
  \usepackage{textcomp} 
\else 
  \usepackage{unicode-math} 
  \defaultfontfeatures{Scale=MatchLowercase}
  \defaultfontfeatures[\rmfamily]{Ligatures=TeX,Scale=1}
\fi
\usepackage{lmodern}
\ifPDFTeX\else
\fi
\IfFileExists{upquote.sty}{\usepackage{upquote}}{}
\IfFileExists{microtype.sty}{
  \usepackage[]{microtype}
  \UseMicrotypeSet[protrusion]{basicmath} 
}{}
\makeatletter
\@ifundefined{KOMAClassName}{
  \IfFileExists{parskip.sty}{%
    \usepackage{parskip}
  }{
    \setlength{\parindent}{0pt}
    \setlength{\parskip}{6pt plus 2pt minus 1pt}}
}{
  \KOMAoptions{parskip=half}}
\makeatother
\usepackage{xcolor}
\usepackage[margin=1in]{geometry}
\usepackage{graphicx}
\makeatletter
\newsavebox\pandoc@box
\newcommand*\pandocbounded[1]{
  \sbox\pandoc@box{#1}%
  \Gscale@div\@tempa{\textheight}{\dimexpr\ht\pandoc@box+\dp\pandoc@box\relax}%
  \Gscale@div\@tempb{\linewidth}{\wd\pandoc@box}%
  \ifdim\@tempb\p@<\@tempa\p@\let\@tempa\@tempb\fi
  \ifdim\@tempa\p@<\p@\scalebox{\@tempa}{\usebox\pandoc@box}%
  \else\usebox{\pandoc@box}%
  \fi%
}
\def\fps@figure{htbp}
\makeatother
\usepackage{float}

\usepackage{bookmark}
\IfFileExists{xurl.sty}{\usepackage{xurl}}{} 
\hypersetup{
  pdftitle={Bayesian sample size calculations for external validation studies of risk prediction models},
  pdfauthor={Mohsen Sadatsafavi\^{}*, Paul Gustafson, Solmaz Setayeshgar, Laure Wynants\^{}\textbackslash dagger, Richard D. Riley\^{}\textbackslash dagger},
  pdfkeywords={Risk prediction; Uncertainty; Sample size; Decision
theory; Bayesian statistics},
  hidelinks,
  pdfcreator={LaTeX via pandoc}}

\title{Bayesian sample size calculations for external validation studies
of risk prediction models}
\author{Mohsen Sadatsafavi\(^*\), Paul Gustafson, Solmaz Setayeshgar,
Laure Wynants\(^\dagger\), Richard D. Riley\(^\dagger\)}
\date{May 2025}

\begin{document}
\maketitle
\begin{abstract}
Contemporary sample size calculations for external validation of risk
prediction models require users to specify fixed values of assumed model
performance metrics alongside target precision levels (e.g., 95\% CI
widths). However, due to the finite samples of previous studies, our
knowledge of true model performance in the target population is
uncertain, and so choosing fixed values represents an incomplete
picture. As well, for net benefit (NB) as a measure of clinical utility,
the relevance of conventional precision-based inference is doubtful. In
this work, we propose a general Bayesian framework for multi-criteria
sample size considerations for prediction models for binary outcomes.
For statistical metrics of performance (e.g., discrimination and
calibration), we propose sample size rules that target desired expected
precision or desired assurance probability that the precision criteria
will be satisfied. For NB, we propose rules based on Optimality
Assurance (the probability that the planned study correctly identifies
the optimal strategy) and Value of Information (VoI) analysis. We
showcase these developments in a case study on the validation of a risk
prediction model for deterioration of hospitalized COVID-19 patients.
Compared to the conventional sample size calculation methods, a Bayesian
approach requires explicit quantification of uncertainty around model
performance, and thereby enables flexible sample size rules based on
expected precision, assurance probabilities, and VoI. In our case study,
calculations based on VoI for NB suggest considerably lower sample sizes
are needed than when focusing on precision of calibration metrics.
\end{abstract}

\let\thefootnote\relax\footnotetext{From Faculty of Medicine and Faculty of Pharmaceutical Sciences, The University of British Columbia (MS); Department of Statistics, The University of British Columbia (PG); British Columbia Centre For Disease Control (SS); Department of Epidemiology, CAPHRI Care and Public Health Research Institute, Maastricht University, and Department of Development and Regeneration, KU Leuven (LW); Department of Applied Health Sciences, School of Health Sciences, College of Medicine and Health, University of Birmingham, and National Institute for Health and Care Research (NIHR) Birmingham Biomedical Research Centre (RDR)}

\let\thefootnote\relax\footnotetext{$\dagger$ Co-senior authors with equal contribution}

\let\thefootnote\relax\footnotetext{* Correspondence to Mohsen Sadatsafavi, Room 4110, Faculty of Pharmaceutical Sciences, 2405 Wesbrook Mall, Vancouver, BC, V6T1Z3, Canada; email: msafavi@mail.ubc.ca}

\section{Introduction}\label{introduction}

Developing risk prediction models or validating existing ones in a new
population represents significant investment in time, resources, and
expertise. Like other empirical experiments, the design of such studies
should be based on objective, transparent, and defendable principles. A
particular aspect of study design is the sample size of such studies.
The field has witnessed significant recent developments on this
front\cite{dobbin2013, riley2021, Riley2021CPMSampleSizeExVal, snell2021, christodoulou2021, Pavlou2021CPMSampleSizeExVal, pate2023, pavlou2024}.
An example is the multi-criteria approach by Riley et al for the sample
size required for external validation studies, targeting pre-specified
widths of the 95\% confidence intervals around metrics of model
performance\cite{riley2021, Riley2021CPMSampleSizeExVal, Riley2024CPMTutorialPart3}.
For binary outcomes, such metrics are related to discrimination,
calibration, and net benefit (NB). The sample size required for each
component is computed separately, with the largest one advertised as the
final requirement\cite{Riley2021CPMSampleSizeExVal}.

In addition to target precision criteria, each component of this
approach requires as input an assumed true value of the metric of
interest in the target population. In reality, we do not know the true
value of model performance metrics with certainty (otherwise there would
be no need to conduct a validation study in the first place). Further,
due to sampling variability, the precision obtained in one particular
dataset may differ substantially from the expected precision for that
sample size. Thus, there is more uncertainty to be accounted for than
the Riley criteria allow.

A Bayesian approach towards sample size determination allows this
uncertainty to be accounted for, which is advantageous on multiple
fronts. First, it enables the full use of existing information on model
performance at the time the study is designed, rather than forcing the
investigator to express their knowledge as the fixed truth. Second, it
enables users to choose from different classes of sample size rules,
namely those that target the expected values of precision targets (as in
Riley criteria), as well as `assurance'-type rules for the probability
of meeting (or exceeding) precision targets. The latter can be
insightful as focusing on expected values alone does not guarantee that
the desired precision will be achieved in one particular dataset, and
the investigator may want a stronger assurance against this. For
example, an investigator that desires an interval width of 0.1 for the
c-statistic might not perceive much benefit from the future interval
being narrower, but might have a strong preference against not meeting
this target. As such, targeting a sample size that will result in, say,
a 90\% probability that the CI width will be \(\le\) 0.1 might be more
appealing than a sample size that targets an expected CI width of 0.1.
Finally, when assessing clinical utility, the relevance of
precision-based criteria is challenged\cite{claxton1999, vickers2023}. A
Bayesian approach enables the use of novel, decision-theoretic
approaches based on Value of Information (VoI) analysis, which focuses
on the expected gain in clinical utility from increasing sample
size\cite{heath2021}. Bayesian approaches for power and sample size
calculations are mostly investigated in the context of experimental
studies that are aimed at interrogating a null
hypothesis\cite{ohagan2005, joseph1997, spiegelhalter1986, adcock1988}.
However, risk model validation studies are not generally
hypothesis-driven. Rather, the focus is on the precise estimation of a
variety of metrics of model performance. As such, Bayesian sample size
considerations in this context are worth exploring.

Hence, in this article, we propose a Bayesian version of the sample size
formula by Riley et al, which enables the investigator to 1) incorporate
their uncertainty around assumed model performance in calculations, 2)
use sample size rules that target assurance probabilities for meeting
chosen criteria, and 3) use rules based on VoI analysis for clinical
utility. The proposed framework can be used in two general ways: to
quantify the anticipated precision or VoI outcomes from a planned study
if the sample size is fixed (as is the case with validation studies that
are based on already collected data), or to determine the minimum sample
size that achieves pre-specified precision or VoI criteria. This
framework can also be used in a hybrid form: determine the sample size
based on certain criteria (e.g., targeting CI width for c-statistic and
calibration slope), and investigate the consequence of the chosen sample
size using other criteria (e.g, assurance probability around NB for the
final sample size). Calculations can be done for the entire sample, as
well as within subgroups, for example, imposing fairness criteria around
the target precision of estimates among minority sub-groups.

The rest of this manuscript is structured as follows. First, we briefly
review common metrics of model performance and multi-criteria sample
size formula by Riley et al\cite{Riley2021CPMSampleSizeExVal}. We will
then introduce our Bayesian extension of this framework based on various
sample size rules. We propose a general approach for characterizing
uncertainty around model performance based on commonly reported
information from previous studies, and outline Monte Carlo sampling
algorithms for drawing from the corresponding distributions. A case
study based on a model for predicting COVID-19 deterioration showcases
the developments. We conclude by suggesting further areas for research.

\section{Methods}\label{methods}

In this section we review the context, common metrics of model
performance, Riley's sample size formulas for external validation
studies, and our proposal for Bayesian sample size determination.

\subsection{Context}\label{context}

We focus on external validation of a model for predicting the risk of a
binary outcome. Broadly speaking, an external validation study is an
endeavor in learning about the joint distribution of predicted risks
(\(\pi\)) from a pre-specified model and observed outcomes (\(Y\): 0
no-event, 1 event) in a target population. The validation sample \(D_N\)
of size \(N\) participants can be seen as \(N\) pairs of predicted risks
and observed results: \(D_N=\{(\pi_i, Y_i)\}_{i=1}^N\).

A validation study does not involve learning about the relationship
between predictors and the outcome. Rather, the focus is on quantifying
the performance of an existing model. For ease of expositions, we assume
one model is being evaluated. This framework can be extended to
multi-model comparisons with relative ease, but we leave this to
subsequent exploration. As is the standard in contemporary practice,
predicted risks for the model are taken as fixed values, representing
the expected outcome probability among all individuals with the same
predictor pattern. The proposed approach is applicable whenever a model
or algorithm provides estimated risks for individuals (regardless of
whether it was developed using regression, machine learning or AI-based
methods), as a validation study is not concerned on how predictors are
used to compute the predicted risk.

\subsection{Common metrics of model
performance}\label{common-metrics-of-model-performance}

Most model development and validation studies report the following
metrics along with associated uncertainty:

\textbf{1) Outcome prevalence:} \(\phi:=\mathbb{E}(Y)\).

\textbf{2) Calibration function} \(h()\): This is the function that
returns the expected actual outcome risk among individuals with a given
predicted risk: \(h(\pi)=P(Y=1|\pi)\). This function converts the
predicted risk for the ith individual, \(\pi_i\), to the true risk for
that person, which we denote by \(p_i\). \(p_i:=h(\pi_i)\), which is the
expected values of observed outcome values across all individuals with
similar predictor patterns. As needed, by \(h\) we refer to parameters
describing \(h()\). Commonly, including in Riley's sample size formula,
\(h()\) is modeled linearly on the logit scale:
\(\mbox{logit}(h(\pi))=\alpha + \beta\mbox{logit}(\pi)\), and thus \(h\)
consists of an intercept (\(\alpha\)) and slope (\(\beta\)). Sometimes
studies report \(\beta\) and `mean calibration', i.e.,
\(\mathbb{E}(Y-\pi)\), or \(\beta\) and observed-to-expected outcome
ratio (\(O/E:=\mathbb{E}(Y)/\mathbb{E}(\pi)\)), but these are equivalent
as for a fixed value of \(\beta\), knowing any of calibration intercept,
mean calibration, or O/E ratio is enough to specify the calibration
line. In other instances, \(h()\) is modeled more flexibly using
non-parametric methods (e.g., based on LOESS
smoothing)\cite{austin2019}. Riley et al, while determining sample size
based on a logit-linear calibration function, suggests visually
inspecting the variability in anticipated smoothed curves once the
sample size is determined\cite{riley2021}.

\textbf{3) c-statistic} \((c)\): This is a measure of the discriminatory
performance of the model, and is the probability of concordance between
the ranking of predicted risks and outcomes among a randomly chosen
pair. Formally, \(c:=P(\pi_2 > \pi_1 | Y_2=1, Y_1=0)\) with
\((\pi_1,Y_1)\) and \((\pi_2,Y_2)\) being two randomly selected pairs of
predicted risks and responses.

\textbf{4) Net benefit (NB)}: Details of NB calculations are provided
elsewhere\cite{Vickers2006DCA}. In a nutshell, at a chosen risk
threshold \(z\), there are at least three treatment strategies
concerning the use of the model: treat no one (with a default NB of 0),
use the model to treat individuals whose predicted risk is
\(\pi \ge z\), or treat all. We index these three strategies by 0, 1,
and 2, respectively. The \(NB()\) function can be written as (for
brevity of notations, in what follows we drop the notation that would
indicate NB-related calculations are based on the chosen risk
threshold):

\[
\label{eq:masterNB}
NB(k)=
\left\{
\begin{array}{lll}
0 & k=0 & \mbox{(treat no one)}
\\ 
\phi se - (1-\phi)(1-sp) \frac {z}{1-z} & k=1 &(\mbox{use model to decide}) 
\\
\phi-(1-\phi) \frac{z}{1-z} & k=2 & (\mbox{treat all})
\\
\end{array}\right.,
\]

where \(se:=P(\pi\ge z|Y=1)\) and \(sp:=P(\pi<z|Y=0)\) are the
sensitivity and specificity of the model at the chosen threshold,
respectively. An important difference between NB and other metrics of
model performance is that NB is a decision-theoretic measure. If using
the model has higher expected NB over other strategies (irrespective of
uncertainties), then it is expected to confer clinical utility and so
the decision would be to use the model. On the contrary, metrics of
model performance such as c-statistic or calibration function do not
consider the decision-making context.

\subsection{Riley's sample size
formulas}\label{rileys-sample-size-formulas}

Citing the commonality of the above-mentioned metrics in risk modeling
studies, Riley et al structured their multi-criteria equations around
desired precision levels (width of 95\%CI) around these
metrics\cite{Riley2021CPMSampleSizeExVal}. In particular, the following
equations were suggested for approximating the standard deviation of the
sampling distribution of the estimator (i.e., standard error {[}SE{]}),
from which Wald-type confidence intervals could be constructed:

\[\mbox{SE}(c)=\sqrt{\frac{ c(1- c)\left[1+(N/2-1)(\frac{1- c}{2- c}) +\frac{(N/2-1) c}{1+ c} \right]}{N^2 \phi(1- \phi)}},\]

\[\mbox{SE}(\log(O/E))=\sqrt{\frac{1- \phi}{N\phi}},\] and

\[\mbox{SE}(\beta)=\sqrt{\frac{I_\alpha}{N(I_\alpha I_\beta-I^2_{\alpha,\beta})}},\]

where \(I_\alpha = {\mathbb{E}}\left( p(1-p) \right)\),
\(I_\beta= {\mathbb{E}}\left(\mbox{logit}(\pi)^2  p(1-p) \right)\), and
\(I_{\alpha,\beta} = {\mathbb{E}}\left(\mbox{logit}(\pi) p(1-p) \right)\),
and \(p=h(\pi)\).

Assuming all measures are of interest, the final sample size is decided
by the largest \(N\) among each component (and by checking whether the
subsequent variability of calibration curves is acceptable - as
explained above). Riley et al proposed also targeting desired CI bands
around standardized NB. However, a Bayesian approach facilitates using
decision-theoretic approaches for NB, addressing the criticisms around
the relevance of conventional inferential methods for
NB\cite{claxton1999, vickers2023}.

Note that the calculations require assumptions about the anticipated
model performance (e.g., c-statistic, calibration slope) and the
distribution of predicted risks in the target population. The authors
suggested using single point-estimates for the true model performance
values, and one chosen distribution for predicted risks. We now propose
a Bayesian approach that enables modeling uncertainties around these
quantities.

\subsection{Going Bayesian}\label{going-bayesian}

A Bayesian approach towards sample size calculation requires 1)
identifying metrics of interest for model performance that will be
estimated from this sample (e.g., c-statistic, calibration slope); 2)
deciding on precision targets for such metrics (e.g., a 95\% CI width of
0.1 for c-statistic); 3) deciding on sample size rules on such metrics
(e.g., targeting the expected CI width, demanding a 90\% assurance that
the CI width will not be larger than targeted, or 90\% assurance that
the future validation study will correctly identify the most optimal
decision at a risk threshold of interest); 4) specifying our current
information about model performance; 5) generating random draws from the
distribution of desired targets that will be processed according to the
specified rules.

In line with the contemporary practice in risk prediction development
and validation, we adopt a mixed Bayesian-likelihood perspective: we use
existing evidence (prior) to quantify our uncertainty about model
performance, but we assume once the future sample is obtained, we will
solely rely on it (the likelihood). We will discuss the implications of
a fully Bayesian approach where the likelihood and prior are combined.

Let \(P_\theta(\pi,Y)\) represent our knowledge about the joint
distribution of predicted risks and outcomes in the target population.
This joint distribution is indexed by parameters \(\theta\), a random
entity summarizing our current knowledge about this joint distribution
(which encompasses our beliefs about model performance in the target
population). We plan to learn about \(\theta\) by collecting a sample of
data \(D_N\) from \(N\) participants in the target population. The key
aspect of a Bayesian approach is that we treat \(\theta\), \(D_N\), and
therefore any quantity (e.g., CI widths) derived from \(D_N\), as random
entities. Implementing this approach involves the following steps: we
specify \(P(\theta)\), our current (prior) information about model
performance in the target population. Repeatedly, we sample from
\(P(\theta)\) and simulate a random validation sample
\(P(D_N | \theta)\), from which performance metrics are quantified and
recorded. Repeating this many times will generate draws from the prior
predictive distributions (henceforth referred to as `pre-posterior'
distribution, the anticipated distribution of the quantity of interest
after the future sample is procured) of these performance metrics for a
validation dataset of a particular sample size. These draws can be
processed according to various sample size rules and assurance
probabilities.

Naturally, Bayesian Monte Carlo sampling results in draws from
pre-posterior distributions of targets (e.g., CI widths) given a fixed
\(N\). Sample size calculation thus becomes a stochastic inverse
problem: find the smallest \(N\) that satisfies a set of sample size
rules. As the latter is detached from the core of a Bayesian workflow,
we primarily focus on calculating precision and VoI metrics for a given
\(N\), but also briefly discuss our implementation of stochastic
root-finding algorithms for sample size determination.

\subsubsection{Sample size rules}\label{sample-size-rules}

Because the Bayesian approach takes estimates of model performance in
the future validation study as random quantities, it invites
establishing sample size rules that take into account such randomness.
For statistical measures of model performance, we consider two sets of
rules: those that target expected precision intervals, and those that
target a probability (assurance) that the anticipated precision will be
at least as good as targeted. For NB as a measure of clinical utility,
we suggest sample size rules that are not precision-based and rather
hone in on our ability in detecting the strategy with the highest NB.

\paragraph{Sample size rules for metrics of discrimination and
calibration}\label{sample-size-rules-for-metrics-of-discrimination-and-calibration}

\(\\\) Broadly speaking, we consider some scalar summary derived by
applying summarizing function \(g()\) to validation data \({D}_N\), to
meet some criterion. In our context, \(g({D}_N)\) returns the 95\%CI
width for the corresponding metric.

\textbf{Expected CI widths (ECIW)}: This is related to the Average
Length Criterion discussed by Joseph et al\cite{joseph1997}. Here, we
target the expected CI width (ECIW) across the distribution of
\({D}_N\):

\[
ECIW(N)=\mathbb{E}_{\mathbf{\theta}}\mathbb{E}_{{D}_N|\mathbf{\theta}}(g({D}_N)).
\]

If the sample size is fixed, \(ECIW(N)\) is reported for each component
as the expected precision of the future study. For sample size
calculation, we seek the minimum \(N\) that results in \(ECIW(N)\)
meeting precision target \(\tau\):
\(\min\{N \in \mathbb{N} | ECIW(N)\le\tau\}\). We derive \(N\)s
separately for each component, and choose the maximum \(N\) as the final
sample size. For example, we might identify the \(N\) required to ensure
that the expected CI widths of the calibration slope and c-statistic all
meet a desired expected CI width.

\textbf{Assurance-type rules based on quantiles of CI width (QCIW)}:
This is related to the modified Worst Outcome Criterion as discussed by
Joseph et al\cite{joseph1997}. Here, we target the probability of
meeting (or exceeding) desired precision targets. For a given sample
size, this approach returns the CI widths corresponding to a desired
quantile \(q\) (e.g., \(q=0.9\) for 90\% assurance):

\[
QCIW(N, q)=F_N^{-1}(q),
\] with \[
F_N(x)=P(g(D_N)\le x)=\mathbb{E}_{\mathbf{\theta}}\mathbb{E}_{{D}_N|\mathbf{\theta}}(I(g({D}_N)\le x))
\] being the CDF of the distribution of the anticipated CI width for
sample size \(N\). Again, for a fixed-\(N\) setup we report \(QCIW(N)\).
For sample size calculation, we find the minimum \(N\) such the the
\(q^{th}\) quantile is not greater than the target CI width \(\tau\):
\(\min\{N \in \mathbb{N} | QCIW(N,q)\le\tau\}\).

\paragraph{Sample size rules for NB}\label{sample-size-rules-for-nb}

\(\\\)\textbf{Optimality Assurance for NB} : This is the probability
that we will \emph{correctly} identify the strategy that has the highest
population NB based on \(D_N\). To proceed, let \(NB_{D_N}(k)\) be the
sample estimate of \(NB(k)\), and \(NB_\theta(k)\) its true value given
\(\theta\). We note that with the future data at hand, the investigator
will declare a winning strategy as the one that has the highest expected
NB solely based on the sample:

\[
W(D_N)=\mbox{arg}\max_k(NB_{D_N}(k)).
\]

Optimality Assurance is the probability that the NB of the strategy that
we will declare as the best in the sample is the maximum possible NB:

\[
Assurance(N)=P\left(NB_\theta(W(D_N)) = \max_k(NB_\theta(k))\right) = \mathbb{E}_\theta\mathbb{E}_{D_N|\theta} I\left(NB_\theta(W(D_N)) = \max_k(NB_\theta(k))\right).
\]

This assurance is in the probability scale that is non-decreasing and
asymptotes to 1 as the sample size is increased.

\textbf{Expected Value of Sample Information (EVSI)}: The key VoI
quantity is \(EVSI(N)\), the expected gain in NB from conducting a
future validation study of size \(N\), compared with the NB of the
decision made with current information:

\[
EVSI(N)=\mathbb{E}_{\theta}\mathbb{E}_{D_N|\theta}\left(NB_{\theta}(W(D_N))\right) - \max_{_k}{\mathbb{E}_\mathbf{\theta} NB_\theta(k)}.
\]

The first term on the right-hand side is the expcted NB of the decision
that we will declare as optimal based on \(D_N\). The second term on the
right-hand side is the expected NB of the best decision under current
information. However, it might be the case that without the validation
study, the model will not be implemented, regardless of its potential
superiority under current information. In which case, this term can be
replaced by the NB of the default strategy (e.g., treating no one or
treating all).

\(EVSI(N)\) is a non-decreasing function. Its maximum value occurs when
\(N\) is infinity; that is, we have access to the entire population and
can unequivocally determine the optimal strategy. The expected gain from
such perfect information is called the Expected Value of Perfect
Information (\(EVPI\)) and can be calculated
as\cite{sadatsafavi2024a, sadatsafavi2022}:

\[    
    EVPI=\mathbb{E}_{\mathbf{\theta}}\max_k(NB_\theta(k)) - \max_k{\mathbb{E}_\mathbf{\theta} NB_\theta(k)}.
\]

EVSI and EVPI can be expressed in true positive or false positive units.
As these units are context-specific, it is more natural to propose a
unit-less metric as a target of sample size rule. We propose `relative'
EVSI (rEVSI) as the ratio of EVSI to EVPI. This value is intuitive and
can be presented as percentage. An \(rEVSI\) of 0.4 for a given sample
size means that the external validation study at this sample size is
expected to reduce the expected NB loss due to uncertainty by 40\%.

Another sample size rule can be based on the Expected Net Benefit of
Sampling (\(ENBS\))\cite{fenwick2020}. \(ENBS\) requires scaling the
EVSI by the expected size of the population affected by the decision,
and subtracting the costs of collecting a sample in the same (true or
false positive) unit as NB. Let \(\mathcal{M}\) be the expected number
of times the decision of interest is to be made, and \(\mathcal{W}\) be
the effort of every additional recruitment for the validation study in
NB units. Then:

\[    
    ENBS(N)=\mathcal{M}EVSI(N)-\mathcal{W} N,
\] and the optimal sample size is the one that maximizes \(ENBS\). This
rule for sample size determination is fully decision-theoretic as it
avoids specifying any arbitrary threshold values. However, it requires
scaling NBs to the population and establishing the trade-off between
sampling efforts and clinical utility; both of these tasks are
context-specific. As such, while we propose this rule here for
completeness, we will not investigate it in the case study.

\subsubsection{\texorpdfstring{Specifying
\(P(\theta)\)}{Specifying P(\textbackslash theta)}}\label{specifying-ptheta}

The Bayesian approach requires specifying \(P(\theta)\), the
distribution of parameters that govern \(P_\theta(\pi,Y)\). This
parameterization can be done in different ways. For example, if a
(pilot) sample form the target population is available, one can adopt a
non-parametric approach based on the Bayesian bootstrap. In this scheme,
\(\theta\) is the vector of weights assigned to each observation in the
pilot sample. There weights are random with a distribution of
\(P(\theta) \sim \mbox{Dirichlet}(1,...,1)\). The empirical distribution
of this weighted sample can be considered as a random draw from the
distribution of the population\cite{rubin1981}. The validation sample
\(D_N\) then can be obtained from sampling with replacement. Details of
such two-level resampling approach is provided
elsewhere\cite{Sadatsafavi2013, sadatsafavi2024a}.

Our focus here is on parametric modeling based on summary statistics
from previous studies. Noting that \(P(\pi,Y)=P(\pi)P(Y|\pi)\), with
\(Y|\pi \sim \mbox{Bernoulli}(h(\pi))\), an intuitive way of
parameterizing \(\theta\) is via specifying \(P(\pi)\), the distribution
of predicted risks, and \(h\), the parameters defining the calibration
function \(h()\). However, specifying the distribution of predicted
risks in the target population is not straightforward, as risk modeling
studies seldom provide such information. Fortunately, specification of
our knowledge in terms of outcome prevalence, c-statistic, and
calibration function (the same components used in Riley's equations),
that is, defining \(\theta=\{\phi, c, h\}\), can, under mild regularity
conditions, fully identify \(P(\pi,Y)\). The regularity conditions are
as follows:

\begin{enumerate}
\def\labelenumi{\arabic{enumi})}
\item
  \(h()\) is monotonically ascending (under the assumption of
  logit-linearity, this is satisfied as long as calibration slope is
  positive), and
\item
  \(P(p)\) is quantile-identifiable; i.e., any two quantiles of the
  distribution are sufficient for uniquely identifying it. Typical
  distributions for risks, including Beta, Logit-normal, and
  Probit-normal satisfy this requirement\cite{sadatsafavi2024}.
\end{enumerate}

The first condition guarantees that the c-statistic relating \(\pi\) to
\(Y\) (which is often reported) is equal to the c-statistic for \(P(p)\)
(as c-statistic is invariant under monotonical transformation of
predictor values). The second condition is a requirement for the
identifiability of \(P(p)\) given its mean (prevalence) and
c-statistic\cite{sadatsafavi2024}.

As an example of this identifiability, consider an outcome prevalence of
0.25, c-statistic of 0.75, calibration slope of 1.1, and O/E ratio of
0.9. Assuming predicted risks have a Logit-normal distribution, the
calibrated risks will also have Logit-normal distribution. The
parameters of the latter are uniquely identifiable from \(\{\phi, c\}\),
which resolves to \(P(p)\sim \mbox{Logitnorm}(-1.3302,1.0395)\) (the
\emph{mcmapper} R package implements our proposed numerical algorithms
for this mapping\cite{mcmapper}). As well, there is a 1:1 mapping
between the O/E ratio and calibration intercept. Given the calibration
slope of of 1.1 and the specified distribution for \(p\), an O/E ratio
of 0.9 uniquely maps to calibration intercept of -0.089 (see footnote of
Table \ref{tab:PREDICT} for additional informaiton). Thus, to generate a
random validation sample, one can sample \(N\) calibrated risk from
\(P(p)\), generate corresponding response values as
\(Y_i \sim \mbox{Bernoulli}(p_i)\) and compute predicted risks as
\(\pi_i=h^{-1}(p_i)\). \((\pi_i,Y_i)\)s created this way will be a
realization of \(D_N\).

Given this identifiability, characterizing our prior information
involves specifying the joint distribution \(P(\phi, c, h)\). Ideally,
the previous analysis would report both the value and an estimated
covariance matrix variance of \(\hat{\theta}\); i.e., there would be
joint inference about the three elements of \(\theta\). These would then
become the prior mean and covariance matrix of the joint prior for
\(\theta\). In typical practice, however, joint inference on such
parameters is not reported, but probability distributions for individual
components can readily be constructed from existing information. For
example, for outcome prevalence, if from a previous study with sample
size of \(n\), \(m\) individuals experience the outcome, our knowledge
can be specified as \(\mbox{Beta}(m,n-m)\). For other components, point
estimates and reported CI bounds can be used, along an assumed
distribution type, to construct distributions. Examples include
specifying a Log-normal distribution for O/E ratio, Normal distribution
for calibration slope, and Beta distribution for c-statistic based on
reported point estimates and bounds of 95\%CI. The accompanying software
provides a flexible way of specifying such distributions, accepting
specification of distribution parameters, moments, or mean and upper
bound of 95\%CI.

Once marginal distributions are constructed, a simple strategy would be
to complete the prior specification by imposing \emph{a priori}
independence between the three components. An optional strategy, which
would be more faithful to the true data generating mechanism, would
involve a parametric bootstrap procedure to recover parameters
interdependence. In this approach, one simulates multiple samples given
\(\theta=\hat{\theta}\), and for each sample records the ensuing
estimates of \(\hat{\theta}\). The empirical correlation matrix of these
simulated estimates can then be taken as the prior correlation matrix,
instead of simply presuming an uncorrelated prior.

This algorithm in itself quantifies the uncertainty for a population
that is exchangeable with the population(s) from which evidence on model
performance is collected. If the evidence is collected from a single
population, this specification assumes model performance is the same
between the source and target populations. On the other hand, if current
evidence is synthesized from multiple populations using meta-analytic
techniques (as in our case study below), this specification assumes the
target population is a random draw from the distribution of the
meta-population. The predictive distribution of model performance in a
new population can thus be used to characterize uncertainties. If the
exchangability assumption does not hold, different steps of this
algorithm can be modified to model population differences. If the
outcome prevalence are expected to be different, one can shift the
distribution of calibrated risks once they are determined in Step
\ref{estimate_p} to match the prevalence in the target population (which
itself should be a random variable indicating our uncertainty about
prevalence). Independently, if evidence is extracted from a model
development study, and there are concerns about the model being
overfitted, one can add a negative penalty term to the calibration slope
in Step \ref{sample} (itself a random variable) representing our
knowledge about the degree of overfitting. A structured examination of
the degree of relatedness between the target population and source
population(s) might help the investigator decide on the need for, and
extent of, modifications\cite{debray2015}.

\subsubsection{Sampling from pre-posterior
distributions}\label{sampling-from-pre-posterior-distributions}

All sample size rules require expectation with respect to the
distribution of parameters (\(\theta\)) and future sample (\(D_N\)).
This naturally invites a Monte Carlo sampling algorithm based on
repeated sampling from \(P(\theta)\), simulating validation samples from
\(P(D_N | \theta)\), and computing the precision targets and VoI metrics
in this sample. Draws from the pre-posterior distributions are then
processed according to the sample size rules of interest (e.g., average
CI widths for the expected CI width criterion, the corresponding
quantile for assurance-based metrics, or VoI for NB).

In addition to this default approach based on simulating \(D_N\), we
propose an approximate two-step approach for CI width targets that does
not involve simulating \(D_N\). This approaches is based on using SE
equations twice. Let \(\theta_k\) be the metric of interest (e.g.,
c-statistic) among \(\theta\)s. First, in the jth iteration of the Monte
Carlo simulation, we specify \(P(\hat \theta_k | \theta)\), the sample
distribution of the metric of interest in the validation sample given
our draw from \(\theta\). This distribution is specified via method of
moments with the first two moments being the corresponding value for
\(\theta_k\) in \(\theta\), and \(SE^2(\theta_k)\) from the relevant SE
equation, which is interpreted in a Bayesian flavor as the distribution
of the future sample estimate. We then obtain a draw from such a
distribution as a realization of \(\hat \theta_k\), which is plugged
into the SE equation again to quantify its CI width.

Table \ref{tab:PREDICT} provides an algorithmic description of both
approaches.

\begin{table}[H]
\caption{Bayesian Monte Carlo algorithm for drawing from the pre-posterior distribution of precision targets \label{tab:PREDICT}}
  \fbox{%
    \begin{minipage}{\textwidth}
      \begin{enumerate}
        \item \label{prev_cstat_type} Assign a distribution to prevalence, c-statistic, and calibration function representing current knowledge (e.g., based on reported point estimates and confidence bands from previous development or validation studies). If calibration function is specified as a line on the logit scale, this can be one of the following.
        \begin{itemize}
          \item Distributions for calibration intercept and slope.
          \item Distributions for O/E ratio and calibration slope$^*$.
          \item Distributions for mean calibration and calibration slope$^*$.
        \end{itemize}
        \item \label{dist_type} Assign a distribution type for calibrated risks $p$ in the source population$^\dagger$
        \item \label{samp_theta} Obtain a sample of size S for $\theta$: $\theta^{(j)}=\{\phi^{(j)}, c^{(j)}, h^{(j)}\}, j=1,2,...,S$.
        Optional: use parametric bootstrapping to induce correlation among $\{\phi, c, h \}$
        \item \label{for_j} For j=1 to S (number of Monte Carlo simulations). 
        \begin{enumerate}
          \item \label{estimate_p} Derive the parameters of $P^{(j)}(p)$, the distribution of calibrated risks in this iteration, given $\phi^{(j)}$ and $c^{(j)}$, given the distribution type assigned in step \ref{dist_type}. 
          \item[] \hspace*{-2\leftmargin} {*Sample-based method} 
          \item \label{sample} Draw $N$ observations for calibrated risks: $p_i^{(j)} \sim P^{(j)}(p), i=1,2,...,N$. Draw corresponding response values $P(Y_i^{(j)}) \sim \mbox{Bernoulli}(p_i^{(j)})$. Calculate the corresponding values of $\pi$: $\pi_i^{(j)}={h^{(j)}}^{-1}(p_i^{(j)})$. Construct ${D_N}^{(j)}=\{(\pi_i^{(j)}, Y_i^{(j)})\}_{i=1}^N$ as the validation sample.
          \item \label{targets} Using ${D_N}^{(j)}$, construct and record precision targets (CI widths): $ciw^{(j)}=g({D_N}^{(j)})$ for each metric of interest. 
          \item[] \hspace*{-2\leftmargin} {*Two-step approach}
          \addtocounter{enumii}{-2}
          \item For any metric $\theta_k$, specify $P(\hat \theta_k | \theta^{(j)})$ using method of moments, with the first moment being true value of $\theta_k$ from $\theta^{(j)}$, and the second moment being $SE^2(\theta_k)$ (and a choice of distribution type$^\#$). Obtain $\hat \theta_k$ as a draw from this distribution. 
          \item Plug $\hat \theta_k$ into the relevant SE equation to compute the precision criterion (CI width). Record this value. 
        \end{enumerate}
        \item[] Next j
        \item Process the draws from the precision targets according the sample size rule specified:\\
          - For expected CI width criteria: $ECIW=\sum_{j=1}^S ciw^{(j)} /S$.\\
          - For quantile (assurance) CI width criteria: $QCIW=\hat{F}_{ciw}^{-1}(q)$, where $\hat{F}_{ciw}$ is the empirical CDF of ciws and $q$ is the desired quantile (e.g., 0.9).
      \end{enumerate}
    \end{minipage}%
  }
  \vspace{0.5em}
  \footnotesize
  $^*${For a given value of prevalence, both these specifications result in a value for $\mathbb{E}(\pi)$, thus requiring finding parameters of $h()$ in $\mathbb{E}(\pi)=\int_0^1 h^{-1}(p)f(p)dp$. For logit-linear calibration function, $h(\pi)=\mbox{expit}(\alpha+\beta \mbox{logit}(\pi))$.  In our implementation, solving for $\alpha$ given $\mathbb{E}(\pi)$ and $\beta$ is programmed as univariate root-finding.}\\   
  $^\dagger${Currently, Beta, Logit-normal, and Probit-normal distributions are modeled in the accompanying R package.}\\
  $^\#${Central limit theorem justifies Normal distribution as the default. This is indeed compatible with the Wald method of constructing CI.}
\end{table}

For VoI analysis, we are not aware of any previous work proposing
Optimality Assurance. As well, the previously proposed algorithms for
validation EVSI were fully Bayesian\cite{sadatsafavi2024a}. Our proposed
algorithm for computing Optimality Assurance and EVSI for the mixed
Bayesian-likelihood setup is presented in Table \ref{tab:VoI}. We note
that given that sample estimates of prevalence, sensitivity, and
specificity all depend on the four frequencies
\(\{N_{tp}, N_{fn}, N_{tn}, N_{fp}\}\) (the number of true positives,
false negatives, true negatives, and false positives, respectively),
\(D_N\) can be fully specified by these four frequencies, which in turn
can be drawn from their respective binomial distributions without
simulating individual-level data.

\begin{center}
  \begin{table}[H]
  \caption {Computation of optimality assurance and EVSI \label{tab:VoI}}
  \noindent\fbox{%
    \parbox{\textwidth}{%
      \begin{enumerate}
        \item[1-4]{Generate $S$ draws $\theta$: $\theta^{(j)}=\{\phi^{(j)}, c^{(j)}, h^{(j)}\}, j=1,2,...,S$ from steps 1-4 of Table \ref{tab:PREDICT}}.
        \setcounter{enumi}{4}
        \item For $j$ = 1 to $S$ (number of Monte Carlo simulations)
        \begin{enumerate}
          \item Derive the parameters of the distribution of calibrated risks given the draws $\phi^{(j)}$ and $c^{(j)}$.
          \item \label{sesp} Calculate true sensitivity and specificity as follows$^*$:
            $$
            \left\{
            \begin{array}{l}
            se^{(j)} = [\int_{h^{(j)}(z)}^1pf^{(j)}(p)dp]/\phi^{(j)}
            \\ 
            sp^{(j)} = [\int_0^{h^{(j)}(z)}(1-p)f^{(j)}(p)dp]/(1-\phi^{(j)})
            \end{array}
            \right.
            $$, where $f^{(j)}()$ is the PDF of the distribution of calibrated risks, and $h^{(j)}()$ is the calibration function, in the $j^{th}$ iteration. 
          \item \label{trueNBs} Calculate true NBs ($NB^{(j)}(k); k\in\{0,1,2\}$) from prevalence, sensitivity, and specificity using equation \ref{eq:masterNB}. Record maximum expected NB: $NBmax^{(j)}=\max_k NB^{(j)}(k)$.
          \item Generate ${D_N}^{(j)}$, defined by $\{N^{(j)}_{tp}, N^{(j)}_{fn}, N^{(j)}_{tn}, N^{(j)}_{fp}\}$ as:\\
            $N^{(j)}_{+}\sim \text{Binomial}(N,\phi^{(j)})$ (number of positive cases in the future sample), \\
            $N^{(j)}_{tp}\sim \text{Binomial}(N^{(j)}_{+},se^{(j)})$,\\
            $N^{(j)}_{fn}=N^{(j)}_{+}-N^{(j)}_{tp}$,\\
            $N^{(j)}_{tn}\sim \text{Binomial}(N-N^{(j)}_{+},sp^{(j)})$,\\
            $N^{(j)}_{fp}=N-N^{(j)}_{+}-N^{(j)}_{tn}$.
          \item \label{sampleNB} Calculate sample esaimtes of NBs: ${NB}_{D_N^{(j)}}(k); k\in\{0,1,2\}$, by plugging in the sample estimates of prevalence, sensitivity, and specificity from ${D_N}^{(j)}$ in equation \ref{eq:masterNB}.
          \item \label{winner} Identify the winning strategy in the sample: $W^{(j)}=\mbox{argmax}_k {NB}_{D_N^{(j)}}(k)$.
          \item \label{winnerNB} Record the true NB of this strategy $NBsample^{(j)}={NB}_\theta(W^{(j)})$.
          \item Record whether the winning strategy had the highest possible NB: $A^{(j)}=I({NB}_\theta(W^{(j)})=NBmax^{(j)})$
        \end{enumerate}
        \noindent Next $j$
        \item Compute the proportion of times the winning strategy has the highest true NB: $Assurance=\sum_{j=1}^S A^{(j)}/S$.
        \item Average NBs from step \ref{trueNBs}: $ENB(k)=\sum_{j=1}^S NB^{(j)}(k)/S$. Pick the strategy that has the maximum ENB: $maxENB=\max_k ENB(k)$. This is the expected NB of the best strategy under current information.
        \item Average $NBmaxs$ from step \ref{trueNBs}: $ENBmax=\sum_{j=1}^SNBmax^{(j)}/S$. From this subtract $maxENB$. This is EVPI.
        \item Average $NBsamples$ from step \ref{winnerNB}: $ENBsample=\sum_{j=1}^S NBsample^{(j)}/S$. From this subtract maxENB. This is EVSI.
      \end{enumerate}
    }%
  }%
  \vspace{0.5em}
  \footnotesize
  $^*${The integrals for sensitivity and specificity require numerical methods (with some exceptions, for example, for the Beta distribution).}
  \end{table}
\end{center}

\subsubsection{Implementation}\label{implementation}

The above algorithms are implemented in the accompanying
\emph{bayespmtools} package\cite{Sadatsafavi2025bayespmtoolsGithub}, in
particular as two functions \emph{bayespm\_prec()} and
\emph{bayespm\_samp()}, for, respectively, computing precision / VoI for
a fixed sample size, or determining the sample size corresponding to a
set of precision / VoI criteria and sample size rules. These functions
expect evidence to be parameterized via four probability distributions,
one for outcome prevalence, c-statistic, calibration slope, and one of
calibration intercept, O/E ratio, or mean calibration. Missing
correlation among these distributions is optionally (active by default)
imputed via the parametric bootstrap method explained earlier. The
correlation is induced to marginal draws for these parameters via the
methods by Iman and Conover\cite{iman1982}. This method offers the
flexibility to use different distribution types for each component (as
opposed to a single multivariate distribution). Both algorithms
currently implement Logit-normal, Probit-normal, or Beta for the
distribution of calibrated risks.

For a specified sample size, \emph{bayespm\_prec()} computes expected CI
widths, quantiles of CI widths, as well as Optimality Assurance and EVSI
using the above-mentioned Monte Carlo sampling algorithms.
\emph{bayespm\_samp()} solves for minimum \(N\) that satisfies any of
the requested precision criteria coupled with a sample size rule (e.g.,
targeting an expected CI width of 0.2 for O/E ratio, or 90\% assurance
that c-statistic CI width\textless0.1). Because for any \(N\), one
iteration of the Monte Carlo simulation generates one draw from the
pre-posterior distribution of CI widths or NBs, finding the minimum
\(N\) that satisfies a given sample size rule is a stochastic
root-finding problem. In our implementation, for targets related to CI
widths (expected value and assurance), we use the Robbins-Monro
algorithm\cite{robbins1951}. This was motivated by the simplicity of
this algorithm which enables simultaneous optimization for all widths
targets. For Optimality Assurance and VoI metrics, the binomial draws
can be vectorized across the entire sample of \(\theta\)s; thus these
calculations can be performed faster than those based on CI widths. We
use the Stochastic-Simultaneous Optimistic Optimization algorithm by
Valko et al\cite{valko2013stochastic} (implemented in the \emph{OOR} R
package\cite{OOR}).

\section{Case study: Predicting deterioration in hospitalized COVID-19
patients}\label{case-study-predicting-deterioration-in-hospitalized-covid-19-patients}

The ISARIC 4C model is a risk prediction model for predicting
deterioration (need for ventilatory support or critical care, or death)
in patients hospitalized due to COVID-19
infection\cite{Gupta2021ISARIC}. The investigators used electronic
hospital records from all 9 health regions of the UK to develop and
validate this model. The London region was left out for external
validation, and the remaining 8 regions were used in internal-external
validation, where one at a time, one region was left out; the model was
fitted using data from all other regions, and its out-of-sample
performance was assessed in the left-out
region\cite{Collins2024CPMTutorialPart1}. Random-effects meta-analysis
was used to pool out-of-sample estimates of discrimination and
calibration metrics. Finally, the investigators used the data from all 8
regions to fit one final model, and evaluated its performance in the
London region.

To use this setup as an informative example, we assume the London region
did not participate in this study, but is now interested in conducting a
validation study of this model in their region. That is, we ignore the
results reported on the performance of the model in the London region,
and consider the internal-external validation results as the information
at hand. We assume that the performance of the model in the London
region will be a random draw from the distribution of performance
observed across other regions (i.e., it is exchangeable with other
regions).

The total development sample size was 70,349 (after removing those with
unknown outcome status), with total number of events (deterioration)
being 30,316, giving rise to an overall outcome prevalence of 0.428.
Pooled estimates of the c-statistic, mean calibration (difference
between average observed and predicted risks), and calibration slope
were 0.76 (95\%CI 0.75--0.77), -0.01 (95\%CI -0.12--0.09), and 0.99
(95\%CI 0.97--1.02), respectively. The research question now is: what
(minimum) sample size is recommended for a new external validation study
in the London region?

\subsubsection{Application of Riley
approach}\label{application-of-riley-approach}

As a baseline comparison, we performed sample size calculations for
validating this model using the multi-criteria method proposed by Riley
et al\cite{Riley2021CPMSampleSizeExVal}. Riley et al used the ISARIC 4C
as an example, and for consistency, we use the same targets: confidence
interval widths of 0.1 for the c-statistic, 0.22 for O/E ratio, and 0.3
for calibration slope. The assumed true values of these metrics were
taken as the summary (pooled) estimates from the above-mentioned
meta-analyses. The results indicate a sample size of 1,056 is required,
which is dictated by the calibration slope. For other components, sample
sizes are as follows: O/E ratio: 425, c-statistic: 359.

\subsubsection{Application of Bayesian
approach}\label{application-of-bayesian-approach}

\textbf{\emph{Specifying}} \(P(\theta)\): Given our assumption about the
exchangeability of populations across UK geographic regions, the
predictive distribution of the model performance in a new region
represents our uncertainty about the model performance in the London
region. However, it is crucial to note that pooled estimates and
confidence bands reported in the original study are for the average
effect and do not represent predictive distributions. To obtain
predictive distributions, we re-performed the meta-analyses pooling
c-statistic, mean calibration (note that while we specify evidence in
terms of mean calibration, as is reported by Gupta et al, we target O/E
ratio for sample size determination), and calibration slope for the 8
regions in the internal-external validation. Gupta et al did not perform
a meta-analysis on prevalence. We therefore performed this additional
meta-analysis using region-specific prevalence data from their report.
Given the large sample sizes, we performed all meta-analyses based on
normality assumption on the original scale of each parameter, aside from
c-statistic, for which a logit-transformation was performed as
recommended by Snell et al\cite{snell2017}. We used method of moments to
derive distribution parameters from the mean and SD of the predictive
distribution for a new population. We specified Beta distribution for
prevalence, Logit-normal for c-statistic, and Normal for mean
calibration and calibration slope. This resulted in the distributions
for the four parameters presented in Table \ref{tab:ISARIC_evidence}.

\begin{table}[H]
\centering
\caption{Evidence synthesis for the ISARIC study, based on predictive distribution of the internal-external validation results$^*$}
\begin{tabular}{|l c c c |} 
 \hline
  Parameter  &   Distribution & Mean(SD) &  95\% Credible Interval$^\dagger$ \\ [0.5ex] 
 \hline\hline
  prevalence & Beta(119.64,159.91) & 0.428 (0.03) & 0.371–0.486 \\ 
  c-statistic & Logit-normal(1.1565,0.0412) & 0.761 (0.006) & 0.746–0.775 \\ 
  mean calibration & Normal(-0.0093,0.1245) & -0.009 (0.125) & -0.253–0.235 \\ 
  calibration slope & Normal(0.9950,0.0237) & 0.995 (0.024) & 0.949–1.042 \\ 
 \hline
\end{tabular}
\label{tab:ISARIC_evidence}
\footnotesize\\
$^*${We extracted study-specific
estimates from digitized figures from the original report which enables us to use three
significant digits.}\\
$^\dagger${Matched to the 95\% prediction interval from the random-effects meta-analysis - see text.}
\end{table}

We did not model between-study correlations (e.g., via a joint
meta-analysis of parameters), after an exploratory analysis (examining
Spearman rank correlation coefficients) did not support the presence of
strong between-study correlation.

\textbf{\emph{Target precision / VoI criteria and sample size rules}}:
For comparability, we target the same CI width as in the Riley approach,
targeting 95\%CI width of 0.22 for O/E ratio, 0.10 for c-statistic, and
0.30 for calibration slopes. For these metrics, we apply both the
expected value and assurance rules. That is, we determine sample sizes
that correspond to the expected CI width being equal to our targets
(this can be interpreted as a direct Bayesian counterpart of
conventional Riley et al's targets). We also demand a 90\% assurance to
meet or exceed the CI width criteria, representing our preference for
obtaining a narrower over wider CI width. As for clinical utility, we
demand a 90\% assurance on NB at the 0.2 threshold; that is, we desire
to be 90\% confident that the strategy that will emerge as the best in
the validation sample is actually the best strategy in the target
population. Finally, we investigate the expected gain in NB for each
component of the sample size on the EVSI curve. This is an example of a
hybrid use of this framework: CI widths and Optimality Assurance are
used as sample size rules, and resulting sample sizes are assessed on in
terms of their EVSI values.

\textbf{\emph{Analysis setup}}: All analyses were performed in
R\cite{R}. We used the \emph{meta} package for random-effects
meta-analysis of internal-external validation results of
ISARIC\cite{meta}. Sample size computations were done using the
\emph{bayespmtools} package, with 10,000 draws from \(P(\theta)\), and
assuming Logit-normal distribution for calibrated risks. We used the
sample-based approach for this analysis.

\textbf{\emph{Results}}: Table \ref{table:SS} provides the sample size
for each component. The expected CI width components are close to their
frequentist counterparts. The assurance-based targets result in slightly
higher sample sizes, as expected. The final sample size of 1,181 is
dictated by the assurance component of the calibration slope.

\begin{table}[H]
\centering
\caption{Sample size components based onconventional Riley approach (top row) and the Bayesian approach (bottom rows)}
\begin{tabular}{|l c c c c |} 
 \hline
 Approach & c-statistic & O/E ratio & calibration slope & NB \\ [0.5ex] 
 \hline\hline
 Conventional Riley approach (expected CI widths) & 359 & 425 & 1056 & 149 \\ 
 Bayesian (expected CI width) & 351 & 430 & 1064 & NA$^*$ \\
 Bayesian (90\% assurance) & 399 & 522 & 1181 & 306 \\
 \hline 
\end{tabular}
\label{table:SS}
\vspace{0.5em}
  \footnotesize\\
  $^*${Not computed as this framework emphasizes computing assurance and VoI metrics for NB.}
\end{table}

\begin{figure}[htbp]
\caption{Expected CI widths (a), 90th quantile of CI width (b), and optimality assurance (c) curves. Shapes on the line pertain to individual sample size components.}
\centering
\begin{tabular}{@{}ccc@{}}
  \includegraphics[width=0.3\linewidth,height=120pt]{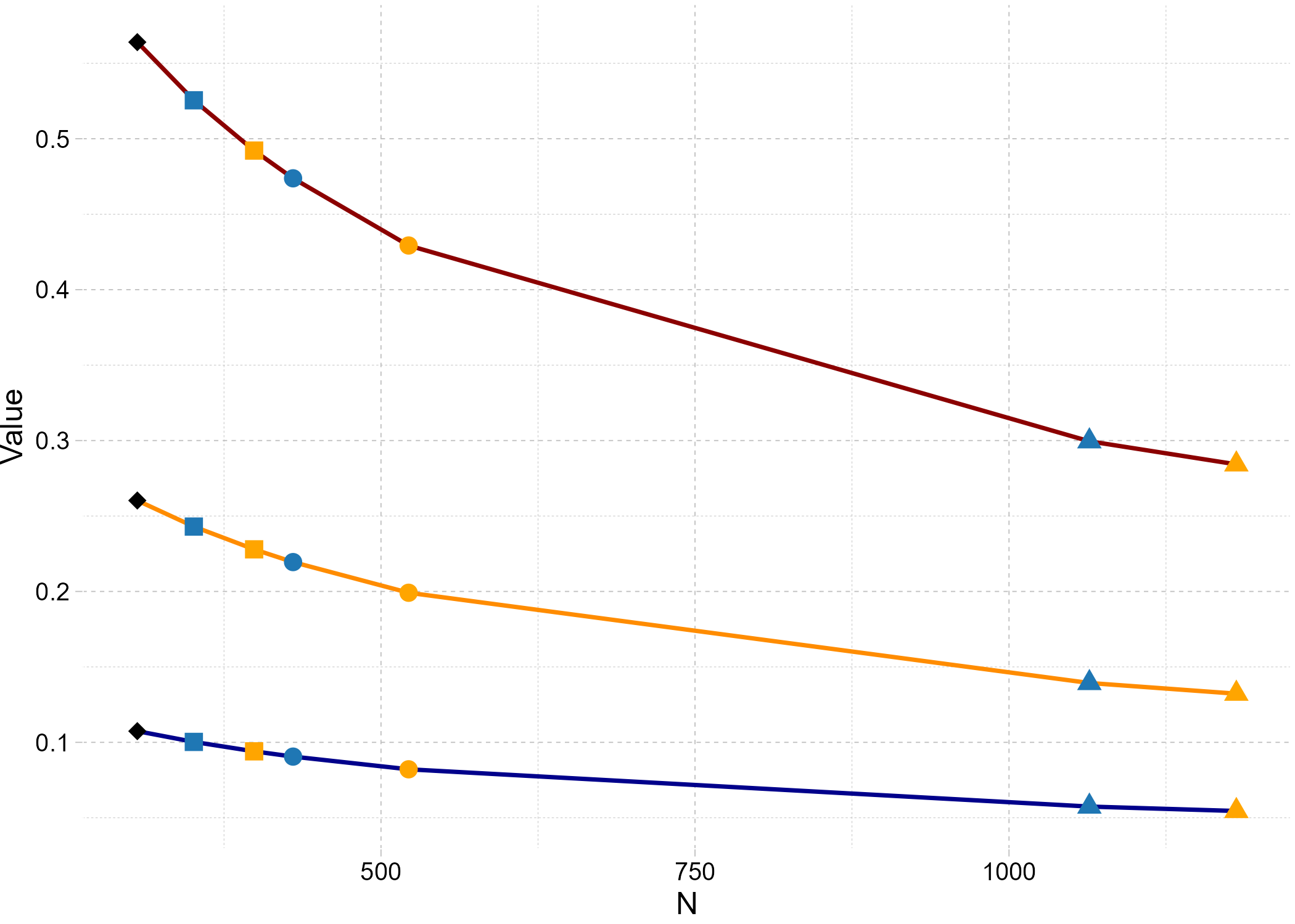} &
  \includegraphics[width=0.3\linewidth,height=120pt]{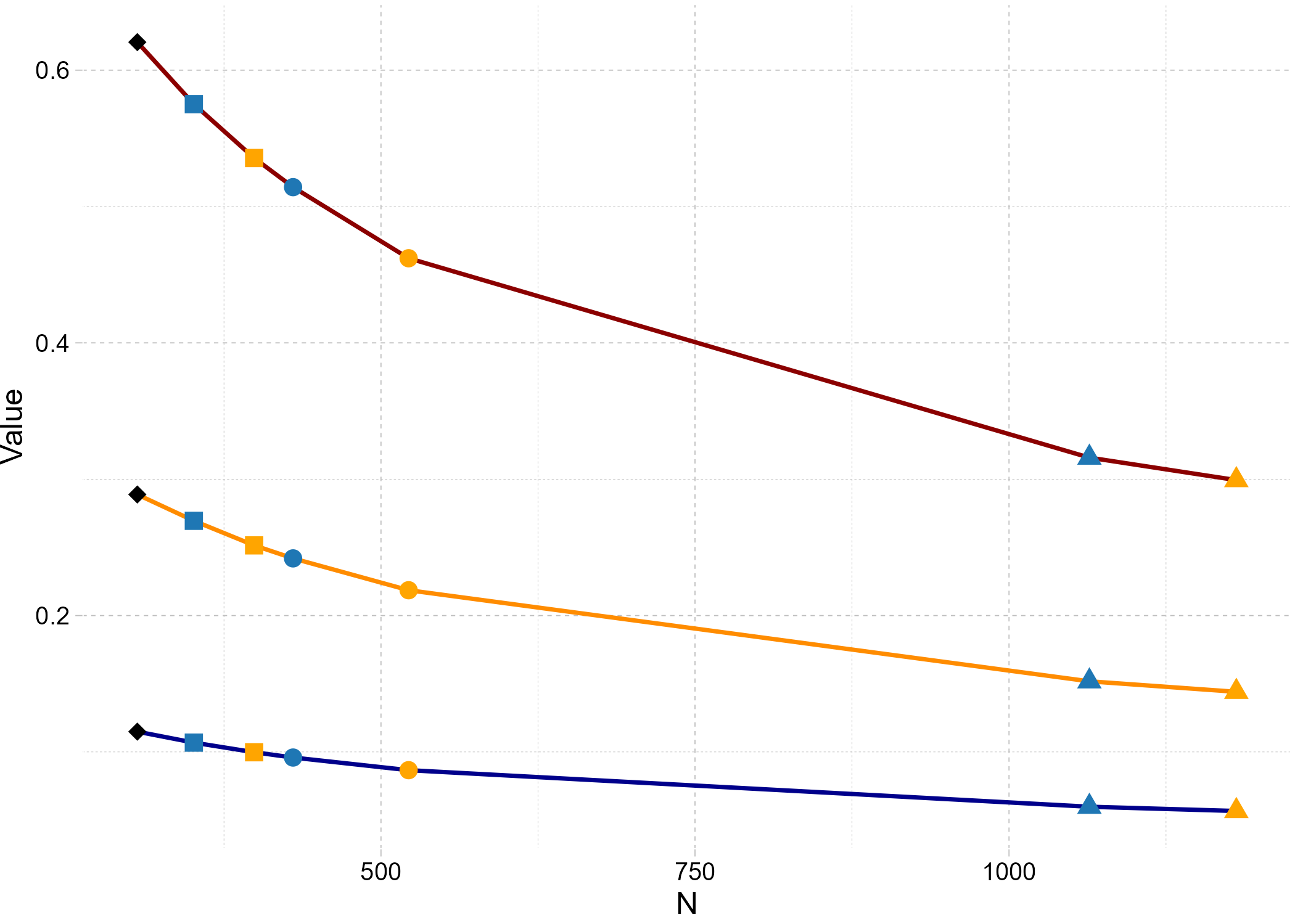} &
  \includegraphics[width=0.3\linewidth,height=120pt]{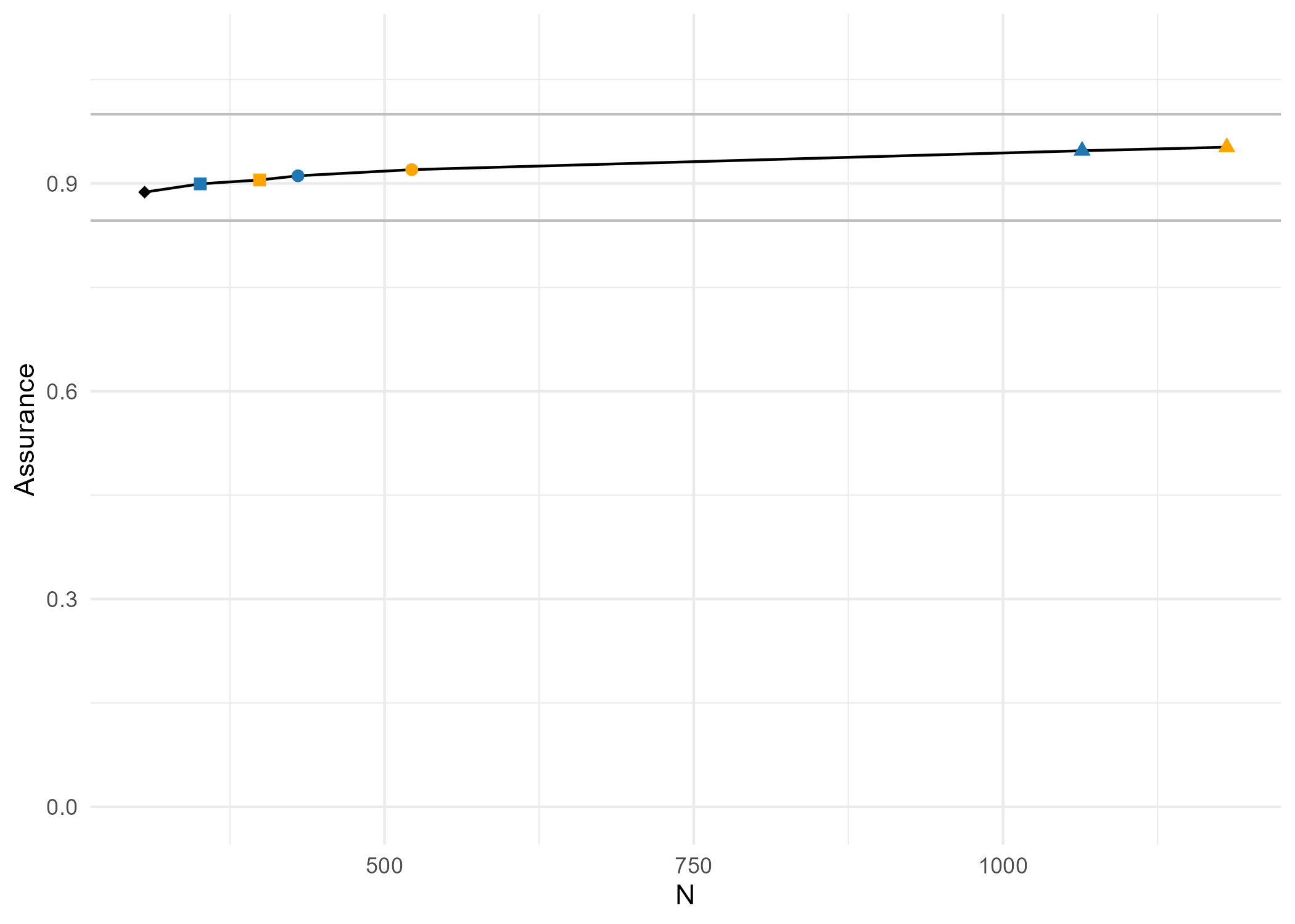} \\
  \multicolumn{3}{c}{
    \includegraphics[width=1\linewidth,height=20pt]{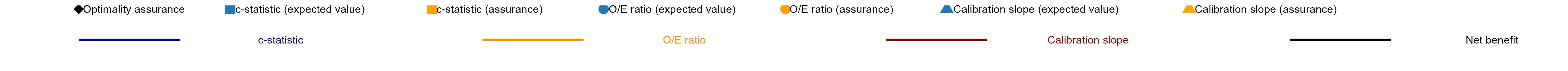}
  } \\
  (a) Expected value of CI widths & (b) 90th percentile of CI widths & (c) Optimality assurance\\[10pt]
\end{tabular}
\label{fig:diagnostics}
\end{figure}

Figure \ref{fig:diagnostics} shows how the precision targets change with
sample size. These are computed from a separate call to
\emph{bayespm\_prec()} (independently of the calculations that
determined the sample sizes) and as such act as diagnostic plots.

\begin{figure}
\caption{Kernel histograms of CI widths for (a) c-statistic , (b) O/E ratio, and (c) calibration slope. Also, (d) shows the kernel histogram of the incremental net benefit of the model compared to the best defualt strategy for two sample sizes}
\centering
\begin{tabular}{@{}ccc@{}}
  \includegraphics[width=160pt,height=120pt]{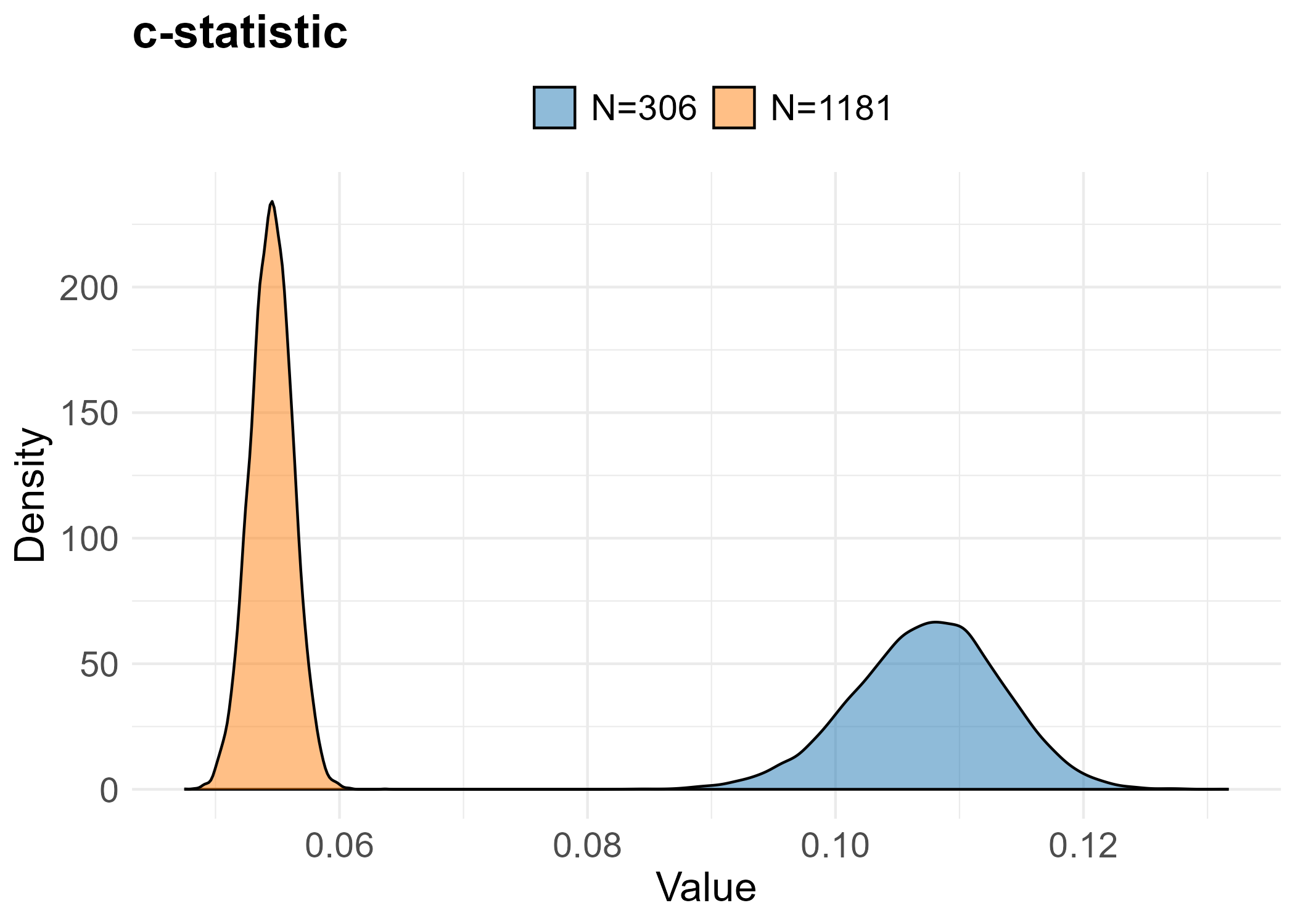} &
  \includegraphics[width=160pt,height=120pt]{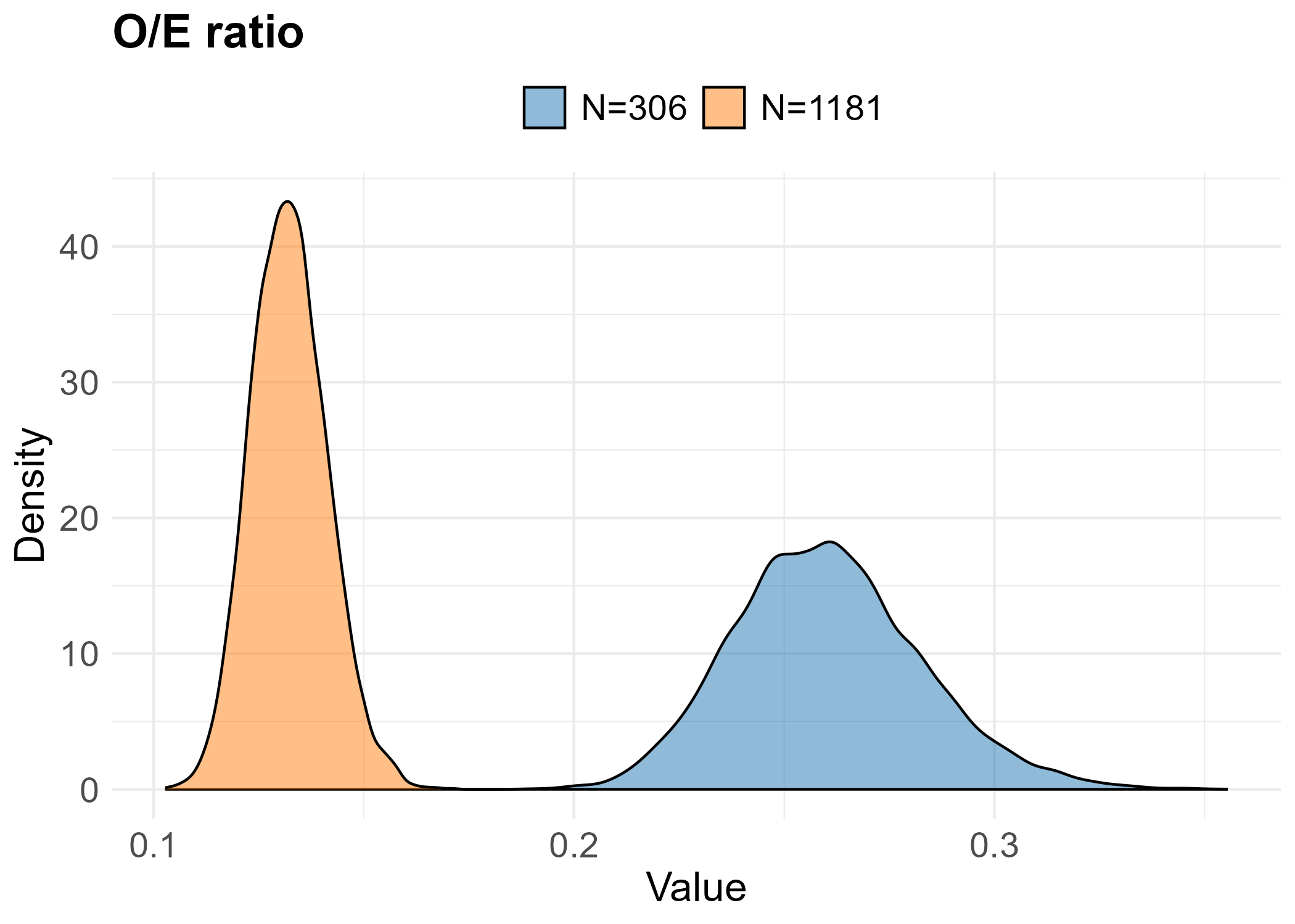} \\
  \includegraphics[width=160pt,height=120pt]{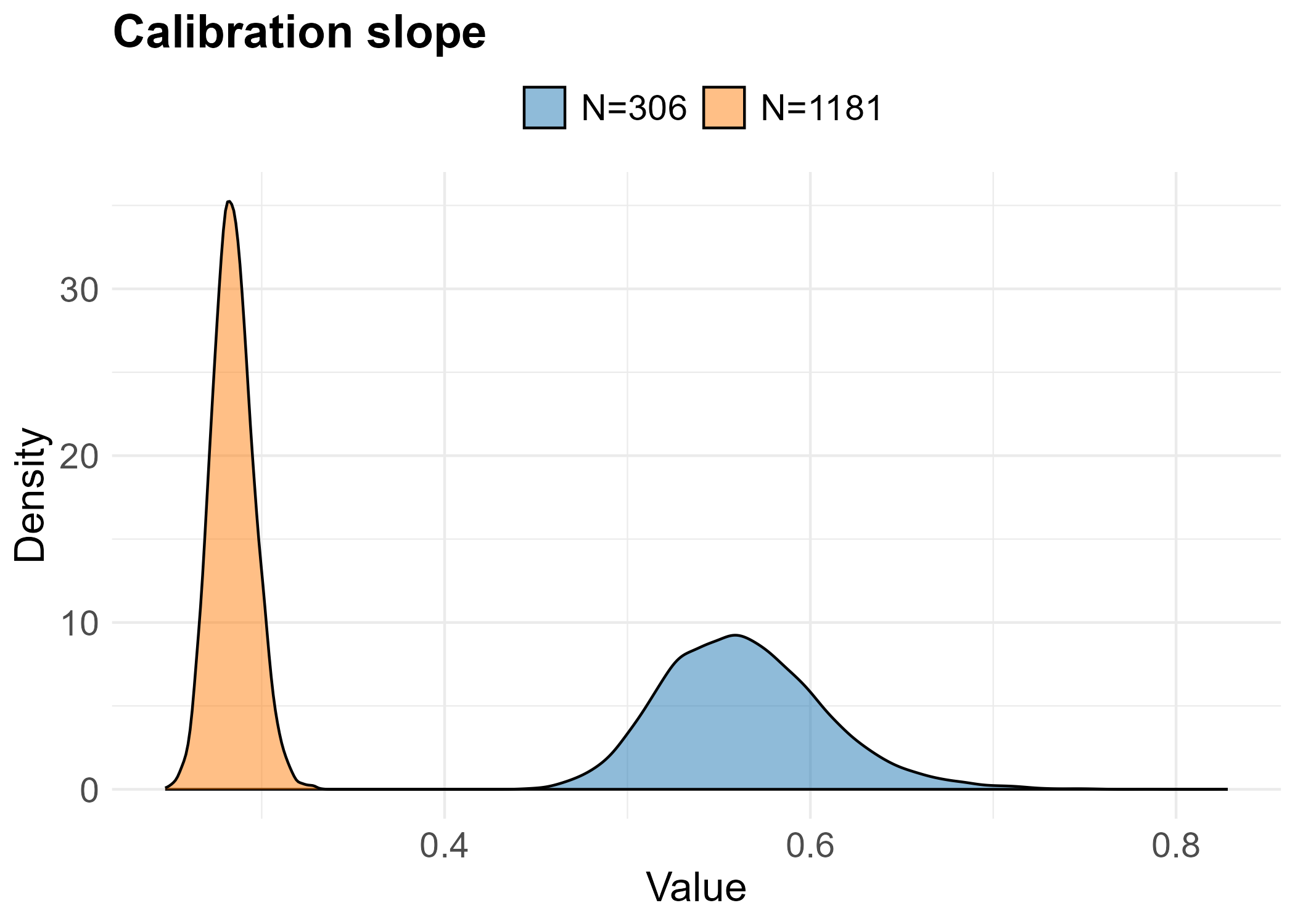} &
  \includegraphics[width=160pt,height=120pt]{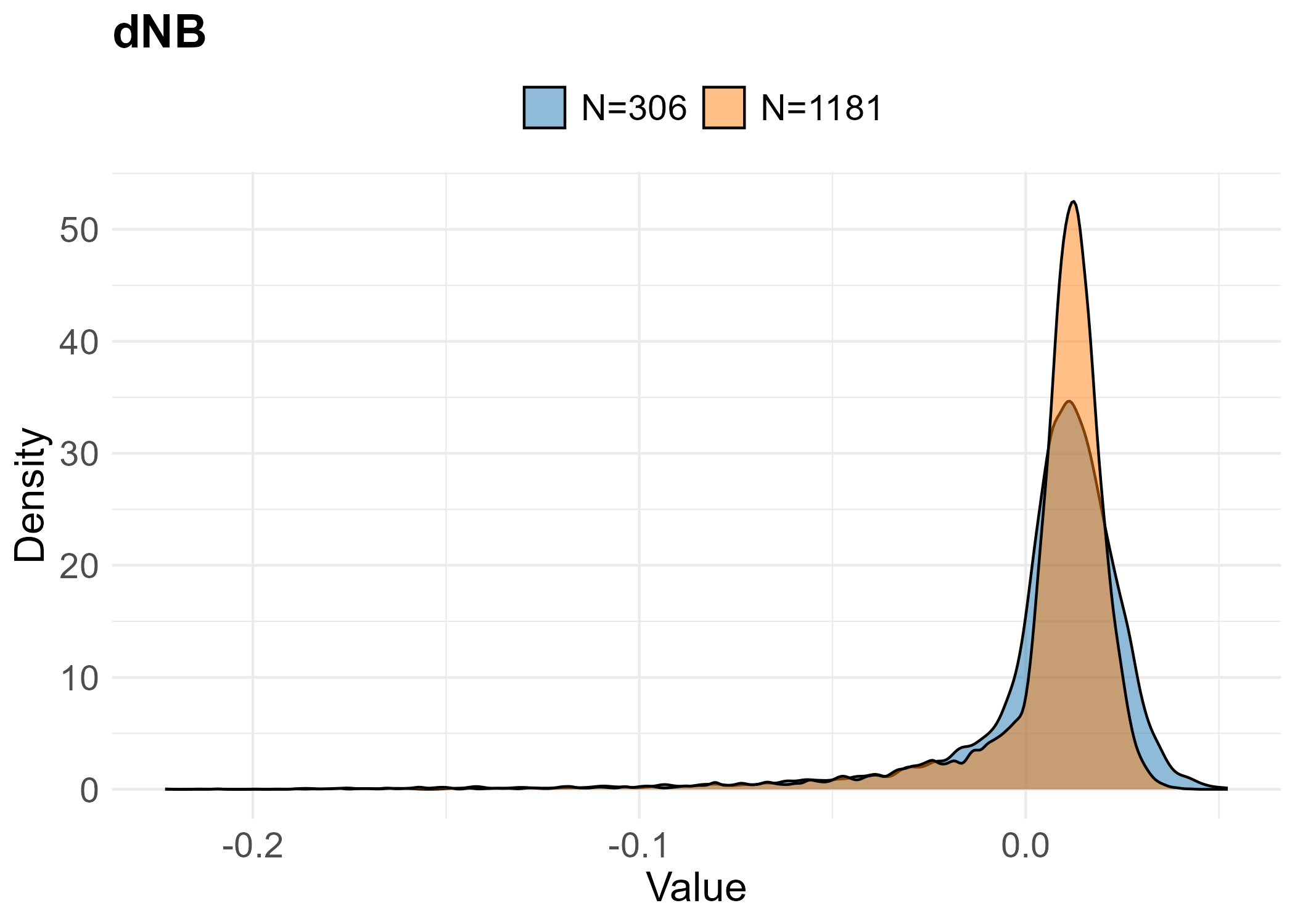} \\
\end{tabular}
\label{fig:hist}
\end{figure}

Figure \ref{fig:hist} shows the exemplary kernel histograms of the
distribution of CI widths (the first three panels) for the smallest
(\(N=306)\) and largest (\(N=1,181\)) components of the sample size. The
last panel demonstrates the distribution of the incremental NB of the
model compared with the default strategies (i.e.,
\(NB_1-\max(NB_0-NB_2)\)). With higher sample sizes, the CI widths get
both shorter and also more clustered. The distributions are relatively
symmetrical. This can explain why the ECIW values are close to their
conventional, frequentist counterparts. For NB, as the sample estimate
of NB is an unbiased estimator, higher sample sizes will result in a
narrower distributions but their location remains the same. For
Optimality Assurance and EVSI, what affects the results is the share of
the distribution that falls on the left side of zero.

\begin{figure}
\caption{Expected Value of Sample Information (EVSI) curve. Shapes pertain to individual sample size components}
\centering
  \begin{tabular}{@{}c@{}}
    \includegraphics [width=1\linewidth]{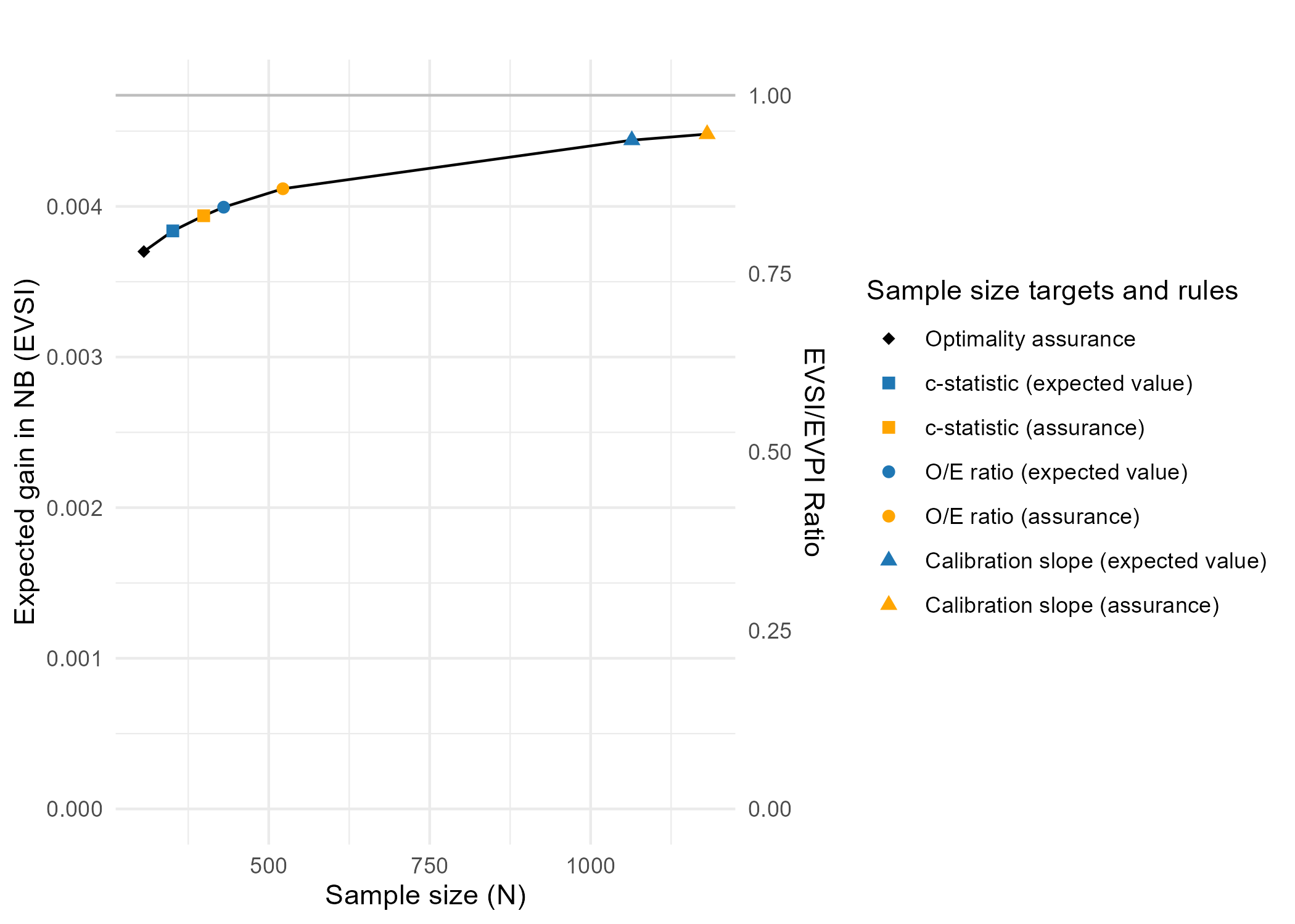} \\ [\abovecaptionskip]
  \end{tabular}
\label{fig:case_study_EVSI}
\end{figure}

Turning to VoI analysis, Figure \ref{fig:case_study_EVSI} provides the
EVSI calculation (with EVSI/EVPI ratio {[}rEVSI{]} as the secondary
Y-axis). Components of sample size calculation are overlaid on the
graph.

From this figure, one can (subjectively) realize that some of the CIW
components fall on the `diminishing return' of the EVSI curve. For
example, going from the the largest sample size based on criteria other
than the calibration slope (N=522) to the sample size dictated by
assurance on calibration (N=1,181), a more than doubling of the sample
size, is associated with only 8.8\% expected gain in NB. This can
justify relaxing the criteria on calibration slope for the validation
study, if the focus is on clinical utility.

\begin{figure}[ht]
    \caption{Bayesian calibration error plots - dashed lines are 95\% credible bands}
    \centering
    \begin{tabular}{cc}
        \includegraphics[width=0.45\textwidth]{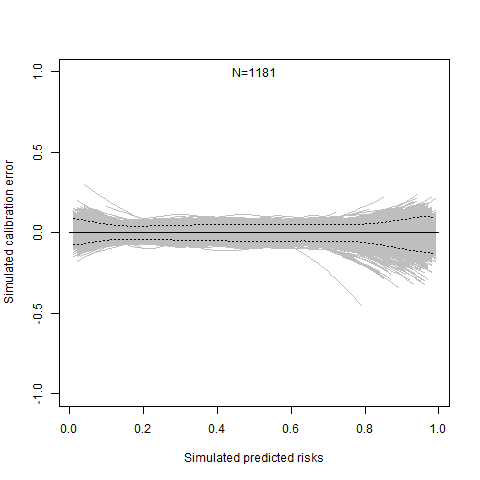}
        \includegraphics[width=0.45\textwidth]{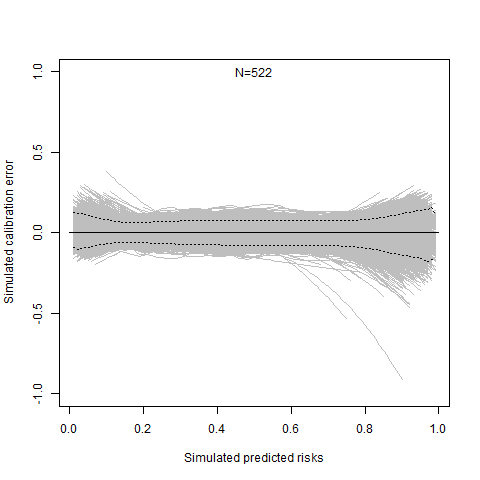} &
    \end{tabular}
    \label{fig:cal_error}
\end{figure}

The decision whether to relax the calibration slope criteria can also be
examined via the stability of the flexible calibration plots. As the
desired precision based on the calibration slope is difficult to
justify, but often dictates the final sample size required, Riley et al
recommend plotting the corresponding variability in calibration curves
that might arise for that sample size\cite{Riley2024CPMTutorialPart3}.
In our context, the variability of a calibration curves has two sources:
the variability in the true calibration function (due to variability in
\(\theta\)) and the variability due to the finite validation sample (due
to variability of \(D_N\)). The former is not a function of sample size.
Because in our Bayesian Monte Carlo sampling algorithm, the true
calibration function is known within each iteration, we can remove
variability due to \(\theta\) to quantify the difference between the
flexible calibration curve and the true calibration function: for each
simulated \(D_N\), we fit a kernel regression (e.g., LOESS) for
\(P(Y=1|\pi)\), then remove \(h(\pi)\) from the fitted values. Figure
\ref{fig:cal_error} shows such Bayesian calibration error plots for the
final sample sizes including calibration slope criteria (N=1,181) and
after removing such criteria (N=522). Moving to the smaller sample size
expectedly increases the spread of calibration errors. Nevertheless,
aside from extreme values of predicted risk, the 95\% credible intervals
(dashed lines) indicate that the error will mostly be \textless0.1.
Again, this can be taken as a further argument in favor of settling on
the smaller sample size. Putting all these together, the final
recommended sample size will be N=522.

\section{Discussion}\label{discussion}

We proposed a Bayesian framework for sample size considerations when a
risk prediction model is to be evaluated (validated) in a target
population. Compared to the conventional frequentist framework, this
approach offers several advantages: it allows investigators to specify
their uncertainties about model performance; in addition to sample size
rules targeting expected CI widths, it enables assurance-type sample
size rules that incorporate investigators' risk preferences; and it
facilitates the the use of Value of Information analysis when clinical
utility of the model is being examined. We proposed characterizing
uncertainties around common metrics of model performance, and introduced
Monte Carlo sampling algorithms for computing precision and VoI metrics.
We implemented this approach as the \emph{bayespmtools} R package, with
two functions, one for quantifying the anticipated precision / VoI
values given a fixed sample size, and one for determining the sample
size given pre-specified rules.

Overall, this framework can provide an objective first step when
designing validation studies aimed at quantifying the performance of a
pre-specified model in terms of trade-off between sample size,
precision, and gain in clinical utility. This framework can be used to
investigate the anticipated precision or expected gain in clinical
utility when the sample size is fixed (based on already collected data).
It can also be used to determine the sample size when original data
collection is planned. A hybrid use, determining the sample size based
on certain criteria and examining the resulting precision and VoI for
other outcomes, is also possible. Our case study demonstrated such a
hybrid approach. We defined rules targeting expected value and 90\%
assurance on the widths of confidence intervals around metrics of
discrimination and calibration, as well as 90\% assurance that we will
be able to correctly identify the strategy with the highest clinical
utility. Then, based on the results of the VoI analysis (as well as
visual inspection of calibration errors), we could argue that one could
potentially relax one component of the samples size requirement
(precision around calibration slope), without losing much precision or
clinical utility.

Among the outputs of this Bayesian approach, the assurance-type rules
and VoI metrics are particularly relevant for risk prediction modeling.
If the estimates of model performance are too uncertain, the adoption of
the model might be met with resistance, even if the point estimates are
compatible with acceptable performance. This, combined with the general
risk-averse attitude of public funding agencies\cite{gross2021}, means
the costs of not meeting the desired targets is typically more than the
benefits of exceeding those targets. Assurance-type rules can be used
alongside those based on expected values to assuage concerns that the
planned study might generate ambiguous results. Further, VoI metrics
address longstanding criticisms on the irrelevance of inferential
statistics when deciding on whether to adopt a health technology
(including markers and risk scoring tools) for patient
care\cite{claxton1999}. These criticism have been recently voiced in the
particular case of NB calculations for risk prediction models
\cite{vickers2023}. Accordingly, we recommend sample size considerations
around NB to move away from confidence intervals and focus on assurance
and EVSI. In particular, EVSI combines the risk of failing to identify
the optimal strategy with the consequences (NB losses) of such a
failure, providing a fuller picture of the implications of learning from
an empirical study in terms of clinical utility. This provides a
`value-based' perspective that can complement the precision-based
approach for sample size determination around metrics of discrimination
and calibration.

There are several areas for further inquiry. We focused on binary
outcomes, but this methodology can be extended to other outcome types.
The application to survival outcomes seems particularly relevant and
feasible. For such outcomes, model performance needs to be assessed at a
time point of interest\cite{McLernon2023}. The time-dependent
equivalents of prevalence and c-statistic are, respectively, the
complement of survival probability and time-dependent area under the
Receiver Operating Characteristic curve\cite{heagerty2000}. If
uncertainty around these metrics are specified, our identifiability
conditions can then be employed to map such metrics to the distribution
of calibrated risks at this time point. However, the impact of censored
observations in the future sample and the presence of competing risks on
precision metrics and VoI values remains to be investigated. We focused
on precision and VoI targets that pertain to the population-level
performance of the model. Sample size consideration can also be
approached in terms of uncertainty at the individual level, such as
instability at individual-level predictions\cite{riley2024}. Further,
when contemplating a validation study, often a secondary objective is to
update (revise) the model if its performance turns out to be
sub-optimal. It makes sense to consider targets related to model
updating when designing validation studies\cite{vergouwe2017}. In line
with Riley equations, we modeled the calibration function to be linear
on the logit scale. This can be relaxed, for example by introducing a
quadratic term as long as one can assure monotonicity of \(h()\) and
specify the joint distribution of the three parameters that would define
it. A more appealing extension would be to model \(h()\)
non-parametrically. This can be done based a non-parametric summaries of
calibration curves, such as those based on kernel smoothing (e.g.,
\(ICI\)\cite{austin2019}) or cumulative plots (e.g.,
\(C^*\)\cite{sadatsafavi2024c}).

From an implementation perspective, Bayesian calculations are inherently
more complex than their frequentist counterparts. Several components of
this framework require numerical integration and optimization methods.
Examples include finding the parameters of \(P(p)\) given prevalence and
c-statistic, mapping O/E ratio or calibration mean to calibration
intercept, and computing sensitivity and specificity (for NB
calculations) at the threshold of interest given \(P(p)\) and \(h()\).
These numerical algorithms might struggle for extreme cases. In
addition, as the Bayesian Monte Carlo sampling produces draws from the
pre-posterior distributions of precision targets (e.g., CI widths),
sample size determination requires stochastic optimization techniques.
These algorithms demand convergence assessment and if needed repetition
with different starting values. While our implementation performed
robustly in our examinations (including the case study), numerical
accuracy and stability of these algorithms should be considered across a
range of input values in dedicated studies. We consider the accompanying
software to be an evolving implementation, subject to revisions and
improvements.

Compared to the conventional sample size determination methods, the
Bayesian approach is predicated on one crucial extra step:
characterizing our uncertainties about model performance. Such
characterization is not currently a common practice in applied
prediction modeling. The conventional wisdom in risk modeling studies is
to report a `final model' as a deterministic function that maps patient
characteristics to an exact value of conditional outcome risk. This
practice ignores the fact that predictions for a model trained using a
finite sample are inherently uncertain. Differences across populations
in the relationship between predictors and the outcome, differences in
predictor assessment approaches across settings, and variations in
outcome ascertainment methods further add to our
uncertainties\cite{Steingrimsson2023}. Fortunately, there have been
recent calls on the importance of characterizing and communicating
uncertainty around model predictions\cite{Riley2025}. These calls
encourage investigators to fully document and communicate uncertainty in
model coefficients in reports of model development (e.g., reporting the
covariance matrix of model coefficients, or reporting the coefficients
of models fitted in bootstrapped copies of the original
sample)\cite{Riley2025}. These can further facilitate incorporating
uncertainties into the design of subsequent studies.

A thought-provoking consequence of adopting a Bayesian approach is the
ultimate fate of the prior information. Reflecting current practices, we
assumed the future validation sample will be the sole source of evidence
on model performance once it is procured, in effect discarding our
current knowledge once it is used to construct \(P(\theta)\). But, if
our existing knowledge is informative enough to warrant investigating
the model in a new setting, should it not be used alongside the sample
to inform our final judgment? Taking our case study as an example, the
information on prevalence alone (based on the moments of its
distribution) is equal to having learned its value from a sample of 281
individuals - a non-trivial amount of information! Incorporating such
prior knowledge into what we learn from the data will reduce prediction
uncertainty, which can affect both patient care (more precise
predictions) and the design of empirical studies (requiring smaller
samples to achieve the same targets). Of course, a perceived drawback is
the subjectivity inherent in real-world Bayesian reasoning. Our case
study might be seen as a stylized setup where predictive distributions
from a meta-analysis were available and the assumption of exchangability
of populations seemed tenable. Things can be more subjective, for
example, when a single development study in another population is the
sole source of evidence. Our current response to the fear of compromised
objectivity is to completely drop the existing information. Perhaps one
can instead define objectivity in terms of explicit and transparent
specification of prior evidence and steps taken to account for
population heterogeneity.

Overall, a cultural shift towards embracing uncertainty in predictions
may also encourage the adoption of study designs that more effectively
leverage existing knowledge. We expect Bayesian approaches towards study
design to gain traction as awareness grows regarding the relevance of
prediction uncertainty in decision making.

\newpage

\bibliographystyle{unsrt}
\bibliography{references}

\end{document}